\documentstyle[prl,epsf,aps,floats,amssymb]{revtex}
\input psfig
\input epsf
\input BoxedEPS.tex
\SetRokickiEPSFSpecial
\HideDisplacementBoxes
\newcommand{\pbarp}{\mbox{$p\bar{p}$}}
\def\ETmiss{\mbox{${\hbox{$E$\kern-0.5em\lower-.1ex\hbox{/}\kern+0.15em}}_{\rm T}$}}

\newcommand{\modeta}{\mid \!\! \eta \!\! \mid}

\def\1800{$\sqrt{s}=1800$ GeV}
\def\630{$\sqrt{s}=630$ GeV}
\def\both{$\sqrt{s}=1800$ and $630$ GeV}
\def\etI{E_{T_1}}
\def\etII{E_{T_2}}
\def\itaI{\eta_1}
\def\itaII{\eta_2}
\def\deta{\Delta\eta}
\def\etab{\bar{\eta}}
\def\xq{($x_1$,$x_2$,$Q^2$)}
\def\xx{($x_1$,$x_2$)}
\def\rap{pseudorapidity}
\def\as{\alpha_s}
\def\ap{\alpha_{\rm BFKL}}

\def\cm{c.m.}

\begin{document}
\preprint{D\O\ Paper XXXX}
\title{Probing BFKL Dynamics 
       in the Dijet Cross Section 
       at Large Rapidity Intervals\\
       in $\bbox{p\bar{p}}$ Collisions
       at $\bbox{\sqrt{s}}$ = 1800 and 630 GeV} 
%
\author{                                                                      
B.~Abbott,$^{46}$                                                             
M.~Abolins,$^{43}$                                                            
V.~Abramov,$^{19}$                                                            
B.S.~Acharya,$^{13}$                                                          
D.L.~Adams,$^{53}$                                                            
M.~Adams,$^{30}$                                                              
V.~Akimov,$^{17}$                                                             
G.A.~Alves,$^{2}$                                                             
N.~Amos,$^{42}$                                                               
E.W.~Anderson,$^{35}$
% ...
\\                                                                            
\vskip 0.30cm                                                                 
\centerline{(D\O\ Collaboration)}                                             
\vskip 0.30cm                                                                 
}                                                                             
\address{                                                                     
\centerline{$^{1}$Universidad de Buenos Aires, Buenos Aires, Argentina}       
\centerline{$^{2}$LAFEX, Centro Brasileiro de Pesquisas F{\'\i}sicas,         
                  Rio de Janeiro, Brazil}                                     
\centerline{$^{3}$Universidade do Estado do Rio de Janeiro,                   
                  Rio de Janeiro, Brazil}                                     
\centerline{$^{4}$Institute of High Energy Physics, Beijing,                  
                  People's Republic of China}                                 
\centerline{$^{5}$Universidad de los Andes, Bogot\'{a}, Colombia}             
\centerline{$^{6}$Universidad San Francisco de Quito, Quito, Ecuador}         
\centerline{$^{7}$Institut des Sciences Nucl\'eaires, IN2P3-CNRS,             
                  Universite de Grenoble 1, Grenoble, France}                 
\centerline{$^{8}$CPPM, IN2P3-CNRS, Universit\'e de la M\'editerran\'ee,      
                  Marseille, France}                                          
\centerline{$^{9}$LPNHE, Universit\'es Paris VI and VII, IN2P3-CNRS,          
                  Paris, France}                                              
\centerline{$^{10}$DAPNIA/Service de Physique des Particules, CEA, Saclay,    
                  France}                                                     
\centerline{$^{11}$Panjab University, Chandigarh, India}                      
\centerline{$^{12}$Delhi University, Delhi, India}                            
\centerline{$^{13}$Tata Institute of Fundamental Research, Mumbai, India}     
\centerline{$^{14}$Seoul National University, Seoul, Korea}                   
\centerline{$^{15}$CINVESTAV, Mexico City, Mexico}                            
\centerline{$^{16}$Institute of Nuclear Physics, Krak\'ow, Poland}            
\centerline{$^{17}$Institute for Theoretical and Experimental Physics,        
                   Moscow, Russia}                                            
\centerline{$^{18}$Moscow State University, Moscow, Russia}                   
\centerline{$^{19}$Institute for High Energy Physics, Protvino, Russia}       
\centerline{$^{20}$Lancaster University, Lancaster, United Kingdom}           
\centerline{$^{21}$University of Arizona, Tucson, Arizona 85721}              
\centerline{$^{22}$Lawrence Berkeley National Laboratory and University of    
                   California, Berkeley, California 94720}                    
\centerline{$^{23}$University of California, Davis, California 95616}         
\centerline{$^{24}$California State University, Fresno, California 93740}     
\centerline{$^{25}$University of California, Irvine, California 92697}        
\centerline{$^{26}$University of California, Riverside, California 92521}     
\centerline{$^{27}$Florida State University, Tallahassee, Florida 32306}      
\centerline{$^{28}$University of Hawaii, Honolulu, Hawaii 96822}              
\centerline{$^{29}$Fermi National Accelerator Laboratory, Batavia,            
                   Illinois 60510}                                            
\centerline{$^{30}$University of Illinois at Chicago, Chicago,                
                   Illinois 60607}                                            
\centerline{$^{31}$Northern Illinois University, DeKalb, Illinois 60115}      
\centerline{$^{32}$Northwestern University, Evanston, Illinois 60208}         
\centerline{$^{33}$Indiana University, Bloomington, Indiana 47405}            
\centerline{$^{34}$University of Notre Dame, Notre Dame, Indiana 46556}       
\centerline{$^{35}$Iowa State University, Ames, Iowa 50011}                   
\centerline{$^{36}$University of Kansas, Lawrence, Kansas 66045}              
\centerline{$^{37}$Kansas State University, Manhattan, Kansas 66506}          
\centerline{$^{38}$Louisiana Tech University, Ruston, Louisiana 71272}        
\centerline{$^{39}$University of Maryland, College Park, Maryland 20742}      
\centerline{$^{40}$Boston University, Boston, Massachusetts 02215}            
\centerline{$^{41}$Northeastern University, Boston, Massachusetts 02115}      
\centerline{$^{42}$University of Michigan, Ann Arbor, Michigan 48109}         
\centerline{$^{43}$Michigan State University, East Lansing, Michigan 48824}   
\centerline{$^{44}$University of Nebraska, Lincoln, Nebraska 68588}           
\centerline{$^{45}$Columbia University, New York, New York 10027}             
\centerline{$^{46}$New York University, New York, New York 10003}             
\centerline{$^{47}$University of Rochester, Rochester, New York 14627}        
\centerline{$^{48}$State University of New York, Stony Brook,                 
                   New York 11794}                                            
\centerline{$^{49}$Brookhaven National Laboratory, Upton, New York 11973}     
\centerline{$^{50}$Langston University, Langston, Oklahoma 73050}             
\centerline{$^{51}$University of Oklahoma, Norman, Oklahoma 73019}            
\centerline{$^{52}$Brown University, Providence, Rhode Island 02912}          
\centerline{$^{53}$University of Texas, Arlington, Texas 76019}               
\centerline{$^{54}$Texas A\&M University, College Station, Texas 77843}       
\centerline{$^{55}$Rice University, Houston, Texas 77005}                     
\centerline{$^{56}$University of Washington, Seattle, Washington 98195}       
}                                                                             
\date{\today}
\maketitle

\begin{abstract}
Inclusive dijet production at large pseudorapidity intervals
($\deta$) between the two jets 
has been suggested as a regime for observing BFKL dynamics.
We have measured the dijet cross section for large $\deta$
in \pbarp\ collisions
at \both\ using the D\O\ detector.
The partonic cross section increases strongly with the 
size of $\deta$. 
The observed growth is even stronger than 
expected on the basis of BFKL resummation 
in the leading logarithmic approximation.
The growth of the partonic cross section
can be accommodated with an effective BFKL intercept of 
$\ap (20 \, {\rm GeV}) = 1.65 \pm 0.07$. 
\end{abstract}


\twocolumn




Jet production in
the high-energy limit of Quantum Chromodynamics (QCD),
as defined by
center-of-mass (\cm) energies ($\sqrt{s}$)
much larger than the
momentum transfers ($Q$), 
presents a very interesting and yet little explored area.
In this kinematic region,
the significantly different energy scales of the process
lead to calculated
jet cross sections characterized by the appearance of
large logarithms $\ln(s/Q^2)$, which must be summed to all 
orders in $\as$.
This summation is accomplished through the
Balitsky-Fadin-Kuraev-Lipatov \mbox{(BFKL)}~\cite{BFKL} equation,
which involves a space-like chain of an infinite number of
gluon emissions. 
The gluons have similar transverse momenta, 
but they are strongly ordered in their pseudorapidities
or, equivalently,
in their longitudinal momentum fractions, $x_i$.
Thus, the BFKL equation effectively describes the evolution in $x$
(growth with $1/x$)
of the gluon momentum distribution in the proton.

Attempts to isolate and probe the BFKL evolution in the low-$x$ region
in $ep$ collisions at HERA, by measuring the forward jet and particle
cross sections~\cite{hera}, have led to ambiguous results.
(At HERA, forward denotes the region away from the current jet and 
 towards the proton remnant.)
In \pbarp\ collisions, the azimuthal decorrelation
as a function of the pseudorapidity interval, $\deta$, in dijet systems
has been studied~\cite{decorrelation}.
(Here, $\eta = -\ln[\tan(\theta/2)]$, 
 where $\theta$ is the polar angle of the jet relative to 
 the proton beam.)
It has been argued~\cite{dokshitzer}, however, 
that the azimuthal angle distribution
is not an inclusive enough quantity for the cancellation between the
real and virtual soft gluons that accompany dijet production; thus,
such a distribution can not be used as a probe of the BFKL equation.

Inclusive dijet production at large pseudorapidity intervals
in high energy \pbarp\ collisions, on the other hand,
provides an excellent testing ground for BFKL dynamics.
We present a measurement of
the dijet cross section at large $\deta$
using the D\O\ detector at the Fermilab Tevatron collider.
We reconstruct the event kinematics using the most forward/backward
jets, 
and measure the cross section as a function of $x_1$, $x_2$ and $Q^2$.
The longitudinal momentum fractions
of the proton and antiproton, $x_1$ and $x_2$,
carried by the two interacting partons are defined as:
\begin{equation}
x_{1,2} = \frac{2 E_{T_{1,2}} }{\sqrt{s}} \; e^{ \pm \etab} \;
                                        \cosh(\deta /2) \;,
\label{eq:x12}
\end{equation}
where $\etI$($\etII$) and $\itaI$($\itaII$) are the 
transverse energy and pseudorapidity of the most forward(backward) jet,
$\deta = \itaI - \itaII \geq 0$,
and $\etab=(\eta_1+\eta_2)/2$. 
The momentum transfer during the hard scattering is defined as:
\begin{equation}
Q = \sqrt{\etI \etII} \;.
\label{eq:q2}
\end{equation}

The total dijet cross section, $\sigma$, can be factorized into the
partonic cross section, $\hat{\sigma}$, and the parton distribution
functions (PDF), $P(x_{1,2},Q^2)$, in the proton and antiproton:
$\sigma = x_1P(x_1,Q^2) \, x_2P(x_2,Q^2) \, \hat{\sigma}$.
The \mbox{partonic} \cm\ energy, $\sqrt{\hat{s}}$, 
equals $\sqrt{x_1 x_2 s}$.
For sufficiently large values of $x_1$ and $x_2$,
any large $\as\ln(s/Q^2)$ terms in $\sigma$
correspond to large $\as\ln(\hat{s}/Q^2)$,
which are of the order of $\as\deta$,
and factorize in $\hat{\sigma}$.
Using the BFKL prescription to sum the leading logarithmic terms
$\as\ln(\hat{s}/Q^2)$ to all orders in $\as$, results in 
an exponential rise of 
$\hat{\sigma}$ with $\deta$~\cite{MN}:
\begin{equation}
\hat{\sigma}_{\rm BFKL} \propto
                  \frac{1}{Q^2} \cdot
                  \frac{e^{(\ap-1)\deta}}{\sqrt{\alpha_s \deta}} \;,
\label{eq:sigma_bfkl}
\end{equation}
where $\ap$ is the BFKL intercept that governs the strength of the 
growth of the gluon distribution at \mbox{small} $x$.
In the leading logarithmic approximation (LLA), $\ap$ is given 
by~\cite{BFKL}:
\begin{equation}
\ap - 1 = \frac{\as(Q) \, 12 \ln2 }{\pi} \;.
\label{eq:ap}
\end{equation}

The predicted rise of the partonic cross section with $\deta$
is difficult to observe experimentally due to the dependence of
the total cross section on the PDF.
To \mbox{overcome} this difficulty,
we measure the cross section at two \cm\ energies,
$\sqrt{s_A}=1800$ GeV and $\sqrt{s_B}=630$ GeV,
and take their ratio 
for the same values of $x_1$, $x_2$ and $Q^2$.
This eliminates the dependence on the PDF, 
and reduces the ratio to that of the partonic cross sections.
The latter is purely a function of the $\deta$ values:
\begin{equation}
R \equiv
    \frac{\sigma(\sqrt{s_A})}
         {\sigma(\sqrt{s_B})}
  = \frac{\hat{\sigma}(\deta_A)}
         {\hat{\sigma}(\deta_B)}
  = \frac{e^{(\ap-1)(\deta_A-\deta_B)}}{\sqrt{\deta_A/\deta_B}} \;.
\label{eq:ratio}
\end{equation}
Thus, varying $\sqrt{s}$,
while keeping $x_1$, $x_2$ and $Q^2$ fixed,
is equivalent to varying $\deta$,
which directly probes the \mbox{BFKL} dynamics.
In addition, measurement of the ratio leads to cancellation of certain 
experimental uncertainties, and enables an
experimental extraction of $\ap$. 

In the D\O~\cite{d0} detector, jets are identified
using the uranium/liquid-argon calorimeters.
These cover the range of $\modeta \leq 4.1$, and are
segmented into towers of $\Delta\eta\times\Delta\phi = 0.1 \times 0.1$
($\phi$ is the azimuthal angle).

The data samples for this analysis were collected during the 1995--1996
Tevatron Collider run.
Events were selected online by a three-level trigger system
culminating in the software trigger requirement of
a jet candidate with 
$E_T > 12$ GeV.
The trigger was 85\% efficient for jets with 
$E_T=20$ GeV, and fully efficient for jets with $E_T>30$ GeV.
The integrated luminosity of the trigger
was 0.7 nb$^{-1}$ for the \1800\ sample,
and 31.8 nb$^{-1}$ for the \630\ sample~\cite{lumi}.

Jets were reconstructed offline using an iterative fixed-cone 
algorithm with a cone radius of ${\cal R}=0.7$ in $(\eta,\phi)$ 
\mbox{space}~\cite{jet-algo}.
The pseudorapidity of each jet was corrected for small 
reconstruction and jet algorithm biases. 
The transverse energy of each jet was corrected in three stages: 
(i) Energy originating from spectator parton interactions, additional
\pbarp\ interactions, noise from uranium decay, and residual energy
from previous \pbarp\ interactions was subtracted on average
from the measured jet energy~\cite{escale};
(ii) The jet energy was corrected for the hadronic response 
of the calorimeter~\cite{escale}; 
(iii) The fraction of the \mbox{particle} energy that showered outside 
of the jet reconstru\-ction cone was recovered, and the fraction of the 
energy reconstructed within the cone that did 
not belong to the original particle was subtracted~\cite{showering}.
The average correction for jets of $E_T\!=\!20$ GeV
and $\modeta = \! 2.5$ is $(22.8\pm4.8)\%$ at \1800;
for jets of the same $E_T$ and $\modeta = 1.2$ the correction is
$(14.5\pm4.0)\%$ at 630 GeV.

The event vertex was required to lie within 50 cm of the detector 
center; 93\%(86)\% of the events at 1800(630) GeV satisfied this
requirement. 
To remove cosmic ray background, 
the imbalance in the transverse momentum of the event was 
required to be less than 70\% of the leading jet $E_T$;
more than 98\% of the events at each \cm\ energy
satisfied this requirement.
To ensure good jet reconstruction efficiency and jet energy calibration,
jets were selected with
$E_T > 20$ GeV and $\modeta < 3$.
Backgrounds from isolated noisy calorimeter \mbox{cells,} 
accelerator beam
losses, and electromagnetic clusters that mimic jets were eliminated
by applying a series of jet quality criteria;
97\% of the jets 
survived this final selection.

The selected jets of each event were ordered in \rap.
A minimum \rap\ interval of $\deta > 2$ was required
between the most forward and most backward jet.
In the final samples, the most forward and most backward jets 
were found to have approximately the same $E_T$.
The values of $x_1$, $x_2$ and $Q^2$ were calculated
from Eqs. (\ref{eq:x12}) and (\ref{eq:q2}).
Most of the data at \1800\ are within $0.01<x_{1,2}<0.30$,
and at 630 GeV, within $0.03<x_{1,2}<0.60$.
The region of maximum overlap, $0.06<x_{1,2}<0.30$, was divided
into six equal bins of $x_1$ and $x_2$.
Due to limited statistics, only one bin in $Q^2$ was used:
$400 < Q^2 < 1000$ GeV$^2$.
The dijet cross section,
corrected for trigger, event and jet selection inefficiencies,
was computed in each \xq\ bin.


The dijet cross section at low \xx\ is affected by the acceptance
of the $E_T>20$ GeV and $\deta>2$ requirements.
To avoid this bias,
we require $x_1 \cdot x_2 > 0.01$.
Similarly, the cross section at high \xx\ is biased by the
$\modeta<3$ requirement, so that we require $x_{1,2} < 0.22$.
A total of ten \xx\ bins satisfy both requirements.

Multiple \pbarp\ interactions during the same beam crossing,
which, in principle, could distort the topology of the event and 
bias the cross section,
were infrequent for the low instantaneous luminosity 
(${\cal L}<10^{30}(2 \times 10^{30})$ cm$^{-2}$s$^{-1}$ 
 at $\sqrt{s} = 1800(630)$ GeV)
data used in this analysis.
\mbox{Nevertheless,} any possible luminosity effects on the
dijet cross section were evaluated by 
measuring the cross section at 630 GeV from lower- and 
higher-luminosity subsamples. No significant difference was
observed between the two measurements.

The dijet cross section is distorted by jet energy resolution.
The resolution was measured as a function of jet \rap\ and $E_T$,
by balancing $E_T$ in events with only two jets back-to-back in $\phi$.
For jets of $E_T=20$ GeV,
the fractional $E_T$ resolution
is 27\%(14\%) at $\modeta \, = 1.2(2.5)$ and \1800
(measured from the jet data collected during the 
 1994-1995 92 pb$^{-1}$ Tevatron run).
At \630, limited statistics prohibited the measurement of the
resolutions in the whole $E_T$ and $\eta$ spectrum.
In the regions where the measurement was possible,
the resolutions at 630 GeV were found to be smaller than 
the resolutions at 1800 GeV by $\sim$1\%.

The distortion of the cross section was corrected using the
{\sc herwig}~\cite{herwig} Monte Carlo (MC) event generator, 
convoluted with the 
\mbox{CTEQ4M}~\cite{cteq4m} PDF.
In the MC events, 
the jet transverse energies were smeared using
the resolutions extracted from the 1800 GeV data.
The $E_T>20$ GeV, $\modeta <3$ and $\deta>2$
requirements were applied separately to the
original fully-fragmented (particle-level) jets and 
to the $E_T$-smeared jets.
Particle-level and smeared dijet cross sections were calculated
in the same \xq\ \mbox{bins} as in the data.
Apart from normalization differences, 
the smeared \mbox{\sc herwig}
cross section at both \cm\ energies
exhibits the same dependence on $x_{1,2}$ as the data.
The ratio of the particle-level to the smeared MC cross 
section in each bin was used as an unsmearing factor to correct
the data cross section for the jet energy resolution \mbox{effects.}
The unsmearing correction for the dijet cross section is typically
of the order of 10\% 
at both \cm\ energies,
whereas the unsmearing correction for the ratio of the cross sections
amounts to only 6\%.
The difference between the measured resolutions at the two \cm\
energies was accounted for in the systematic uncertainties.
The unsmearing method was verified by using a smeared MC sample
generated with {\sc isajet}~\cite{isajet},
and comparing the {\sc isajet} particle-level cross section
to that obtained using our unsmearing procedure based on {\sc herwig}.


The dijet cross sections for $\deta>2$ at \both\
in the selected \xx\ bins are shown in Table~\ref{tab:ratio-ap}.
In each bin, the average values of $x_1$, $x_2$ and $Q^2$
are in good agreement, within the precision of our measurement,
between the \mbox{two} \cm\ energies.
This ensures the cancellation of the PDF
in the ratio of the cross sections.
Also shown in the Table are the values for the BFKL intercept, $\ap$,
extracted from the cross sections and the average
pseudorapidity intervals at 1800 and 630 GeV
in each \xx\ bin,
using Eq.~(\ref{eq:ratio}).

The mean value of the ratios of the cross sections in the ten bins
is equal to
$\langle R \rangle \equiv \langle \sigma_{1800}/\sigma_{630} \rangle
= 2.8 \pm 0.3 \: {\rm (stat)}$.
The mean value of $\ap$
is equal to
$1.65 \pm 0.05 \: {\rm (stat)}$.
The mean pseudorapidity interval, $\langle \deta \rangle$,
in the selected bins
is equal to 4.6 units at 1800 GeV and 2.4 units at 630 GeV.

\begin{table}[htbp]
\begin{center}
\caption{The dijet cross sections for $\deta>2$ at \both\
         and the extracted value of the BFKL intercept
         in each of the ten \xx\ bins.
         The minimum jet $E_T$ is 20 GeV.
         The uncertainties are statistical.}
\begin{tabular}{cclcl}
$x_1$ range &$x_2$ range & $\;\;\;\;\sigma_{1800}$
                         & $\sigma_{630}$
                         & $\;\;\;\;\ap$
                                                             \\
            &            & $\;\;\;\;\;$(nb)
                         & (nb)
                         &
                                                       \\ \hline\hline
 0.06--0.10 & 0.18--0.22 & $28.1\pm6.9$
                         & $8.4\pm0.9$
                         & $1.74\pm0.13$
                                                             \\ \hline
 0.10--0.14 & 0.14--0.18 & $40.1\pm9.5$
                         & $8.8\pm0.9$
                         & $1.83\pm0.11$
                                                             \\
            & 0.18--0.22 & $\;\:3.6\;^{+\;\;4.1}_{-\;\;2.3}$
                         & $5.4\pm0.6$
                         & $0.96\;^{+\;\;0.49}_{-\;\;0.28}$
                                                             \\ \hline
            & 0.10--0.14 & $27.9\pm7.3$
                         & $8.4\pm0.8$
                         & $1.71\pm0.13$
                                                             \\
 0.14--0.18 & 0.14--0.18 & $10.4\;^{+\;\;6.1}_{-\;\;5.0}$
                         & $5.0\pm0.6$
                         & $1.50\;^{+\;\;0.29}_{-\;\;0.24}$
                                                             \\
            & 0.18--0.22 & $\;\:5.6\;^{+\;\;4.5}_{-\;\;3.8}$
                         & $2.9\pm0.5$
                         & $1.44\;^{+\;\;0.38}_{-\;\;0.32}$
                                                             \\ \hline
            & 0.06--0.10 & $26.3\pm6.6$
                         & $8.6\pm0.9$
                         & $1.71\pm0.14$
                                                             \\
 0.18--0.22 & 0.10--0.14 & $12.5\;^{+\;\;6.3}_{-\;\;5.4}$
                         & $6.3\pm0.7$
                         & $1.46\;^{+\;\;0.24}_{-\;\;0.21}$
                                                             \\
            & 0.14--0.18 & $\;\:6.8\;^{+\;\;5.0}_{-\;\;3.2}$
                         & $3.1\pm0.4$
                         & $1.50\;^{+\;\;0.34}_{-\;\;0.23}$
                                                             \\
            & 0.18--0.22 & $\;\:2.4\;^{+\;\;2.8}_{-\;\;1.7}$
                         & $1.7\pm0.3$
                         & $1.28\;^{+\;\;0.60}_{-\;\;0.37}$
                                                             \\
\end{tabular}
\label{tab:ratio-ap}
\end{center}
\vspace{-3mm}
\end{table}

The largest sources of systematic uncertainties on the ratio of the 
cross sections and the BFKL intercept are the jet energy scale 
(yielding an 8\% uncertainty on the ratio and 2\% on the intercept)
and the jet energy resolutions
(7\% on the ratio and 2\% on the intercept).
The individual components of these were evaluated for
correlations between the two data samples.
Additional sources of systematic uncertainties on the ratio and the 
intercept include 
the choice of the input PDF in the Monte Carlo used for unsmearing
(1\% on the ratio, negligible on the intercept)
and the uncertainty in the normalization of the luminosity
(2\% on the ratio and 1\% on the intercept).
The total systematic uncertainty amounts to 11\% on the ratio of the
cross sections and 3\% on the BFKL intercept,
yielding the final results:\\
$\langle R \rangle 
  = 2.8 \pm 0.3 \: {\rm (stat)} \pm 0.3 \: {\rm (sys)}
  = 2.8 \pm 0.4$,\\
$\langle \ap \rangle
    = 1.65 \pm 0.05 \: {\rm (stat)} \pm 0.05 \: {\rm (sys)}
    = 1.65 \pm 0.07$.
Hence, for the same values of $x_1$, $x_2$ and $Q^2$,
the dijet cross section at large $\deta$
increases by almost a factor of three
between the two \cm\ energies,
corresponding
to the increase of $\langle \deta \rangle$
from 2.4 to 4.6 units.

Several theoretical predictions can be compared to our measurement.
Leading order QCD predicts the ratio of the cross sections
to fall asymptotically toward unity with increasing $\deta$.
For the $\deta$ values relevant to this analysis,
the predicted ratio is
$R_{\rm LO}\!=\!1.2$~\cite{Orr-Stirling}.

The {\sc herwig} MC provides an alternative prediction.
It calculates the exact $2 \rightarrow 2$ subprocess, including
initial and final state radiation and angular ordering of the emitted 
partons.
Using the same \xq\ bins as in the data yields
$R_{\rm HERWIG}   = 1.6 \pm 0.1 \: {\rm (stat)}$.

The LLA BFKL intercept according to Eq.~(\ref{eq:ap}) for
$\as(20\,{\rm GeV})=0.17$~\cite{Orr-Stirling} is
$\alpha_{\rm BFKL,\:LLA} = 1.45$.
For $\deta_{1800}=4.6$ and $\deta_{630}=2.4$,
Eq.~(\ref{eq:ratio}) yields
$R_{\rm BFKL,\:LLA} = 1.9$.
It should be noted, however, that the leading log approximation
may be too simplistic, and that exact quantitative predictions including
the next-to-leading logarithmic ~\cite{BFKL-NLL}
corrections to the BFKL kernel 
are not as yet available.

It is evident that the growth of the dijet cross section with $\deta$
(from $\langle \deta \rangle = 2.4$ to $4.6$)
is stronger in the data than in the theoretical models we considered.
The measured ratio is higher by 4 standard deviations
than the LO prediction, 3 deviations than the {\sc herwig}
prediction, and 2.3 deviations than the LLA BFKL prediction.

It should be noted that the $x_{1,2}$ definitions
of Eq.~(\ref{eq:x12})
have been kept the same in the data and 
in the theoretical calculations.
Modifying these definitions to account for all jets in the event
changes the ratio of the cross sections 
by less than 10\%.

\begin{figure}[htbp]
\vspace{1mm}
\epsfxsize=2.6in
\centerline{
\epsfbox{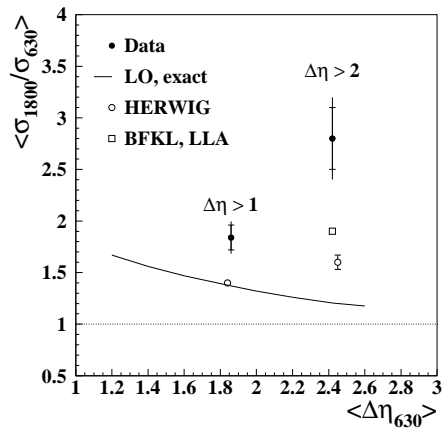}}
\vspace{1mm}
\caption{The ratio of the dijet cross sections at \both\
         for $\deta>1$ and $\deta>2$.
         The minimum jet $E_T$ is 20 GeV.
         The inner error bars on the data points
         represent statistical uncertainties;
         the outer bars represent statistical and uncorrelated
         systematic uncertainties added in quadrature.
         The error bars on the {\sc herwig} predictions
         represent statistical uncertainties.
         The LO and BFKL predictions are analytical calculations.}
\label{fig:ratio_vs_y}
\vspace{1mm}
\end{figure}

Finally, the $\deta>2$ requirement was changed to $\deta>1$,
and the analysis was repeated. 
For $\deta > 1$, Eq. (\ref{eq:x12}) yields 
$x_1 \cdot x_2 > 0.005$,
which results in a selection of fifteen unbiased \xx\ bins.
The mean pseudorapidity interval in the selected bins is equal
to 4.2 at 1800 GeV and 1.9 at 630 GeV.
The average ratio of the 1800 and 630 GeV cross sections 
in the selected bins was measured to be
$1.8 \pm 0.1 \: {\rm (stat)} \pm 0.1 \: {\rm (uncorrelated \: sys)}$.
The results are shown in Fig.~\ref{fig:ratio_vs_y}
as a function of the mean pseudorapidity interval at \630.
In the case of the $\deta>1$ requirement,
the observed ratio is once again
larger than the exact LO and {\sc herwig} \mbox{predictions.}
It is interesting, however, that {\sc herwig} exhibits
the same qualitative behavior as the data in that the ratio
of cross sections decreases as the $\deta$ requirement is relaxed,
whereas the exact LO calculation predicts a very different trend. 
(A BFKL prediction is not shown for the case of $\deta>1$
since the pseudorapidity interval is not sufficiently large for
the formalism to be meaningful.)

In conclusion, we have measured the dijet cross section for large
pseudorapidity intervals 
at \both, 
and the ratio of the cross sections for the same values of $x_1$, $x_2$
and $Q^2$ at the two energies.
The latter corresponds to the ratio of the partonic cross sections
for different values of $\deta$.
The measured partonic cross section increases strongly with $\deta$,
more strongly than expected on the basis of any current prediction.

We appreciate the many fruitful discussions with A.~Mueller,
L.~Orr and J.~Stirling.
We thank the staffs at Fermilab and at collaborating institutions
for contributions to this work, and acknowledge support from the
Department of Energy and National Science Foundation (USA),
Commissariat  \` a L'Energie Atomique and
CNRS/Institut National de Physique Nucl\'eaire et
de Physique des Particules (France),
Ministry for Science and Technology and Ministry for Atomic
   Energy (Russia),
CAPES and CNPq (Brazil),
Departments of Atomic Energy and Science and Education (India),
Colciencias (Colombia),
CONACyT (Mexico),
Ministry of Education and KOSEF (Korea),
CONICET and UBACyT (Argentina),
A.P. Sloan Foundation,
and the Humboldt Foundation.

\end{document}